\documentclass[11pt]{article}

\usepackage[english]{babel}
\usepackage[latin1]{inputenc}
\usepackage{lmodern,csquotes}
\usepackage[T1]{fontenc}

\usepackage{amssymb,amsmath,amsthm,latexsym,amsfonts,amscd,dsfont,enumerate,bbm}
\usepackage{a4wide,graphicx,tikz}

\usepackage{algorithm,algpseudocode}
\usepackage{pgfplots}
\usepackage{subfig}
\usepgfplotslibrary{groupplots}
\usepackage{marginnote}
\usepackage{soul}

\usepackage[backend=biber,style=authoryear-comp,maxcitenames=1,sorting=nyt,sortcites=false]{biblatex}
\definecolor{mylightyellow}{rgb}{1,1,.8}
\definecolor{mylightgreen}{rgb}{.8,1,.8}
\definecolor{mydarkred}{RGB}{178,34,34}
\definecolor{mydarkgreen}{RGB}{34,139,34}
\definecolor{mydarkblue}{RGB}{72,61,139}
\definecolor{mydarkyellow}{RGB}{218,165,32}

\algdef{SE}[SUBALG]{Indent}{EndIndent}{}{\algorithmicend\ }%
\algtext*{Indent}
\algtext*{EndIndent}
\renewcommand{\Comment}[2][.5\linewidth]{%
  \leavevmode\hfill\makebox[#1][l]{//~#2}}

\usepackage[]{hyperref}
 \hypersetup{
 colorlinks=true,breaklinks=true,
 urlcolor= mydarkblue,linkcolor= mydarkblue,citecolor= mydarkgreen,
 pdfauthor={Manzano Herrero, A.P., Nastasi, E., Pallavicini, A., V\'azquez Cend\'on, C.},
}

\newtheorem{theorem}{Theorem}[section]
\newtheorem{remark}[theorem]{Remark}

\newcommand{\ind}[1]{1_{\{#1\}}}

\newcommand{\argmin}[1]{\underset{{#1}}{\operatorname{arg\,min}}\,}

\title{Pricing commodity index options \footnote{The opinions here expressed  are solely those of the authors and do not represent in any way those of their employers.}}
\author{
Alberto Pedro Manzano-Herrero\thanks{ Department of Mathematics and CITIC, University of A Coru{\~{n}}a, Spain {\tt alberto.manzano.herrero@udc.es}}
\and
Emanuele Nastasi\thanks{swissQuant Group AG, Zurich, Switzerland {\tt nastasi@swissquant.com}.}
\and
Andrea Pallavicini\thanks{Intesa SanPaolo Milan, Italy, {\tt andrea.pallavicini@intesasanpaolo.com}.}
\and
Carlos V\'azquez \thanks{Department of Mathematics and CITIC, University of A Coru{\~{n}}a, Spain, {\tt carlos.vazquez.cendon@udc.es}}
}
\date{
\small First Version: September 24, 2021.  This version: \today
}

\begin{document}

\maketitle

\begin{abstract}

We present a stochastic local volatility model for derivative contracts on commodity futures. The aim of the model is to be able to recover the prices of derivative claims both on futures contracts and on indices on futures strategies. Numerical examples for calibration and pricing are provided for the S\&P GSCI Crude Oil excess-return index.

\end{abstract}

\bigskip

\noindent {\bf AMS classification codes:} 65C05, 91G20, 91G60.

\newpage

\maketitle

\pagestyle{myheadings} \markboth{}{{\footnotesize A. Manzano, E. Nastasi, A. Pallavicini \& C. V\'azquez, Pricing commodity index options}}
\tableofcontents
\section{Introduction}
\label{sec:introduction}


Futures are the most liquid contracts in commodity markets, hence they are the main tool to gain exposure on the underlying commodities. As futures expire on a fixed date, if a trader wants to maintain exposure to price changes of the underlying commodity, he/she needs to replace his/her current position on near-term expiring futures with a new position on futures with longer maturities. This strategy of shorting the position with shortest maturity and longing a new position with a longer maturity is called a ``rolling strategy'' or a ``rolling hedge''. There exist specific implementations of rolling strategies, such as the ones addressed in the S\&P GSCI indices\footnote{GSCI refers to Goldman Sachs Commodity Index. The white paper describing the index methodology can be found at the S\&P web site: \url{https://www.spglobal.com/spdji/en/documents/methodologies/methodology-sp-gsci.pdf}.}. Market participants willing to take exposure in the commodities market can trade derivative contracts on an index avoiding the associated burdens of directly trading futures, such as implementing themselves a rolling strategy or dealing with margin procedures. The goal of this article is to propose a model for futures prices able to calibrate plain-vanilla option prices on both commodity futures and indices. Since contracts on indices are usually sensitive to smile effects and include path-dependency, the pricing model must be able to describe both curve and smile dynamics.

It is well-known that local volatility (LV) models, presented in \textcite{dupire1994pricing,Derman1994}, can match exactly the market volatility surface for plain-vanilla options, that is, they can reproduce the volatility smile. However, LV dynamics can be unrealistic, since it produces a flattening in implied forward volatilities as pointed out in  \textcite{rebonato1999volatility}, so that it could be inadequate to price contracts on indices. This flattening of the implied forward volatilities does not occur in Stochastic Volatility (SV) models, where the variance of the asset satisfies its own stochastic differential equation (SDE) such as in the Heston model, see \textcite{heston1993closed}. Nevertheless, since SV models have a parametric form, they cannot be exactly calibrated to market plain-vanilla option prices. As an attempt to obtain a model which can both be exactly calibrated to plain-vanilla options and reproduce the stochastic behaviour of the volatility, Stochastic Local Volatility (SLV) models have been proposed in the literature. SLV models were originated during the late 90s and the early 2000s, see \textcite{said1999pricing,lipton2002masterclass,ren2007calibrating} for example, and nowadays they constitute a standard for pricing in many markets.

If we take a closer look to our commodity settings, we see that in general to describe the dynamics of the futures term structure, we should consider a different SV or SLV model for the price of each futures contract, see for instance \textcite{Pilz2011} and \textcite{chiminello}. Such models are an accurate description of the dynamics of the futures term structure. However, since usually only a limited number of quotes can be found in commodity markets, they might be difficult to calibrate so that we prefer to look for a more parsimonious description. Here, we follow the approach first depicted in \textcite{nastasi2020smile}, where the authors introduce a smaller family of SLV processes to drive the dynamics of the whole futures term structure. 

\medskip

The paper is organized as follows. In Section~\ref{sec:index} we briefly describe the strategy to replicate a position on the index. Then, in Section~\ref{sec:model} we illustrate the modelling framework and we explain the calibration procedure based on Markov projections along with the shortcuts we can take when we are only interested in strategies such as the one given by S\&P GSCI indices. Finally, in Section~\ref{sec:numerics} we investigate some numerical examples of the calibration procedure on West Texas Intermediate (WTI) crude oil futures and plain-vanilla options along with plain-vanilla options on the S\&P GSCI WTI Crude Oil index.

\section{GSCI indices}
\label{sec:index}


S\&P GSCI indices are designed to replicate performances of actual commodity sectors. The idea behind the construction of the S\&P GSCI indices is to create an index that simulates a continuous investment on a basket of commodities (or a single commodity). The calculation of the S\&P GSCI indices takes into account the fact that a person holding positions in a contract near expiration would need to roll such positions forward as they approach settlement or delivery dates. For this reason, the methodology for calculating S\&P GSCI indices includes a rolling procedure designed to replicate the rolling of actual futures positions. Since executing the rolling procedure on a single day could be difficult to implement or, if completed on a single day, could have an adverse impact on the market, such rolling would take place over a period of several days.

As the mechanics of the rolling depends particularly on which specific index we are dealing with, from now on we will focus on S\&P GSCI Excess-Return (ER) indices, which represent the daily return of a portfolio of commodity futures contracts. The rolling forward of the underlying futures contracts occurs once each month, on the fifth through ninth business day (the roll period) and the index is calculated as though these rolls occur at the end of each day during the roll period at the daily settlement prices. In the next subsections we explain the specific strategy of the non-rolling and rolling periods. We will illustrate how the index price is calculated if it depends on a single underlying commodity, which is the interest in this article.

\subsection{Evolution during a non-rolling period}


On any business day during a non-rolling period the value of the S\&P GSCI ER index is equal to the product of the value of the index on the preceding business day times one plus the contract daily return on the business day on which the calculation is made. 

For each futures maturity date $T_i, \,i=1,\ldots,n$, we denote $F_t(T_i)$ the futures price observed at time $t\le T_i$. Moreover, let $I_t$ denote the value of the index at time $t$. The investment strategy implemented by the index consists in buying, at the beginning of each day $t$, a quantity $Q_t$ of futures contracts on the front month such that the nominal value of the investment is exactly $I_t$. Therefore, the amount $Q_t$ of contracts we buy is given by
\begin{equation}
Q_t := \frac{I_t}{F_t(T^c)} \,,
\label{norollingQuantity}
\end{equation}
where $T^c$ indicates the maturity of the front month. Due to the market movement of the futures price, at the end of the day our investment will have generated a profit-and-loss equal to
\begin{equation}
W_{t+1} := Q_t \left( F_{t+1}(T^c) - F_t(T^c) \right) = I_t \left( \frac{F_{t+1}(T^c)}{F_t(T^c)} - 1 \right) \,.
\end{equation}
Such profit-and-loss is invested again in the strategy whose new value becomes
\begin{equation}
I_{t+1} = I_t + W_{t+1} = I_t \,\frac{F_{t+1}(T^c)}{F_t(T^c)} \,.
\end{equation}
If we repeat the same strategy the next day we get that the invested amount remains unchanged
\begin{equation}
Q_{t+1} = \frac{I_{t+1}}{F_{t+1}(T^c)} = \frac{I_t}{F_t(T^c)} \,.
\end{equation}
Proceeding recursively we have that the index value at $n$-th day after $t$ can be calculated directly from the initial conditions as follows
\begin{equation}
I_{t + n} = I_t \,\frac{F_{t+n}(T^c)}{F_t(T^c)} \,.
\label{norollingStrategy}
\end{equation}

\subsection{Evolution during a rolling period}


On a rolling period, we need to roll the nearest futures contract $T^c \equiv T_1$ to the second nearest futures contract $T^f\equiv T_2$ at a rate of 20\% per day for the five days of the roll period. Notice that in other indexes different from the S\&P GSCI ER index the rolling procedure is done between contracts with other maturities (e.g.\ the GSCI Gold index). We could also adapt our methodology to these indices in a straightforward way.

Until just before the end of the fifth business day, the entire S\&P GSCI ER index portfolio consists of the front futures contracts. At the end of the fifth business day, the portfolio is adjusted so that 20\% of the held contracts are in the second futures contract, while 80\% remain in the front one. The roll process continues on the sixth, seventh and eighth business days, with relative weights of front to second contracts of 60\%/40\%, 40\%/60\% and 20\%/80\%. At the end of the ninth business day, the last contract of the old front futures is exchanged, thus completing the roll and leaving the entire portfolio in what we have been calling the second futures contract. At this time, this former second futures becomes the new front contract, and a new second futures is formed (with futures maturities further in the future) for use in the next month roll.

The last key point about the roll process is to specify exactly what the 80\%/20\% or other relative splits between front and second contracts mean. The roll percentages refer to contracts or quantities, not values. Taking the first day of the roll as an example, just before the roll takes place at the end of the day, the S\&P GSCI consists of the front futures contract. That portfolio, constructed the night before and held throughout the fifth business day, has a dollar value. For the roll, that dollar value is distributed across the front and the second futures such that the number of contracts or the quantity of the front ones is 80\% of the total and the quantity held of the second ones is 20\% of the total.

We can illustrate the mechanics of the rolling period with an example to clarify the terms. Starting from the fifth business day of the month, due to the approaching expiry of the front month, the investment is gradually spread between the front and second futures. We call $\alpha(t)$ the investment percentage on the front futures contract  $F_t(T^c)$, and $1-\alpha(t)$ the investment percentage on the second futures contract $F_t(T^f)$. Then, we define $Q_t$ within the rolling period as
\begin{equation}
Q_t := \frac{I_t}{\alpha(t) F_t(T^c) + (1-\alpha(t)) F_t(T^f)} \,,
\label{rollingQuantity}
\end{equation}
which represents the purchased quantity of a fictitious contract made by the combination of the first and second futures. This investment generates the following profit-and-loss
\begin{equation}
W_{t+1} := I_t \frac{\alpha(t) ( F_{t+1}(T^c) - F_t(T^c) ) + (1-\alpha(t)) ( F_{t+1}(T^f) - F_t(T^f) ) }{\alpha(t) F_t(T^c) + (1-\alpha(t)) F_t(T^f)} \,,
\end{equation}
so that the new value of the strategy is given by
\begin{equation}
\label{eq: evolution_roll}
I_{t+1} = I_t + W_{t+1}  = I_t \frac{\alpha(t) F_{t+1}(T^c)  + (1-\alpha(t)) F_{t+1}(T^f) }{\alpha(t) F_t(T^c) + (1-\alpha(t)) F_t(T^f)} \,.
\end{equation}
Contrary to the non-rolling period, if we now consider the evolution to the next day we obtain that the quantities have changed due to the change in the investment weights
\begin{equation}
\begin{split}
Q_{t+1} = \frac{I_t}{\alpha(t) F_t(T^c) + (1-\alpha(t)) F_t(T^f)} \frac{\alpha(t) F_{t+1}(T^c)  + (1-\alpha(t)) F_{t+1}(T^f) }{\alpha(t+1) F_{t+1}(T^c) + (1-\alpha(t+1)) F_{t+1}(T^f) } \,,
\end{split}
\end{equation}
so that the value of the index at time $ t + 2 $ cannot be evaluated only as a function of the initial conditions.

The description of the index strategy leads us to introduce a futures price model able to describe the complex path-dependent dynamics of rolling periods. In the next section we deal with this problem.

\section{Modelling commodity futures}
\label{sec:model}


The prices of plain-vanilla options on a commodity index (such as the ones discussed in Section~\ref{sec:index}) have a direct dependence on the paths followed by futures prices, so that we need to correctly model not only the futures volatility smile quoted in the market, but also its dynamics. As stated in Section~\ref{sec:introduction}, usually a complex term-structure dynamics can be modelled by allowing each futures price to be driven by its own stochastic process, while smile dynamics can be modelled by means of stochastic volatilities. For this reason, we will start by introducing a SLV model for the prices of futures contracts, and we try to do so in a parsimonious way.

\subsection{The modelling framework}


We start by defining the dynamics of the futures prices. The index dynamics will be derived by implementing the index definition given in the previous section in terms of the underlying futures. For each futures maturity date $T_i, \, i=1,\ldots,n$, let $F_t(T_i)$ be the futures price observed at time $t\le T_i$. We describe the dynamics of futures prices under the risk-neutral measure as in~\textcite{nastasi2020smile}, and we consider deterministic interest rates. We assume that each futures price follows a SLV model given by
\begin{equation}
dF_t(T_i) = L(t,T_i,F_t(T_i)) \,\sqrt{v_t} \,dW^i_t,
\label{eqn:SLV_model}
\end{equation}
for $i=1,\ldots,n$, where the leverage functions $\{L(t,T_i,K)\}_{i=1,\ldots,n}$ represent the local-volatility components of each SLV model, and they are assumed to be  Lipschitz, positive and at most of linear growth in price, so that we can guarantee the existence of a solution for each SDE in \eqref{eqn:SLV_model}. The common variance process $v_t$ satisfies the SDE
\begin{equation}
dv_t = \kappa (\theta - v_t) + \chi \sqrt{v_t} \,dW^v_t \,,
\label{eqn:variance}
\end{equation}
where $\kappa$ is the mean-reversion speed of the variance, $\theta$ is the long-term mean variance, and $\chi$ is the volatility of the variance (also known as vol-of-vol). The initial value of the variance $v_0$ is not directly observable, so that we consider it as an additional parameter to be calibrated. For simplicity, all these parameters are assumed to be constant. Moreover, $\{W^i_t\}_{1,\ldots,n}$ and $W^v_t$ are standard Brownian motions under the risk-neutral measure with correlations
\begin{equation}
d\langle W^i,W^j \rangle_t = \rho_{ij} \,dt
\;,\quad
d\langle W^i,W^v \rangle_t = \rho_v \,dt
\end{equation}
for $i,\,j=1,\ldots,n$. Notice that we are assuming a common variance process and the same correlation value between such process and the other Brownian motions since we aim at a parsimonious description of futures dynamics when possible.

The next steps are the specifications of the leverage functions and the correlation structure. In particular we will look at the calibration procedure to understand how to define them in the most simple form, so that we are able to recover the prices of plain-vanilla and index options on futures.

\subsection{The leverage functions}


If we look at the SLV literature, for instance in \textcite{guyon2012being}, we can find the description of a practical procedure to calibrate the leverage function by means of Markov projections via the application of the Gy\"ongy Lemma, see \textcite{gyongy1986mimicking}. This Lemma states under which conditions the marginal densities of two semi-martingales are equivalent in law. Thus, if we are able to calibrate plain-vanilla options quoted by the market by means of a simpler model, e.g.\ a LV model, we can ensure that the SLV model does the same if such models satisfy the hypotheses of the Lemma.

In our case, the market quotes plain-vanilla options of futures prices, so that we introduce a LV model for each futures price, which is given by
\begin{equation}
d{\hat F}_t(T_i) = \eta_F(t,T_i,{\hat F}_t(T_i)) \,d{\hat W}^i_t \,,
\label{eqn:futures_lv_equation}
\end{equation}
where the local-volatility functions $\{\eta(t,T_i,K)\}_{i=1,\ldots,n}$ are assumed to be Lipschitz, positive and at most of linear growth in price, so that there exists a solution for the SDE. Then, we can apply the Gy\"ongy's Lemma to perform the matching of the marginal distributions, and we obtain
\begin{equation}
\eta^2_F(t,T_i,K) = L^2(t,T_i,K) \,\mathbb{E}[v_t|F_t(T_i)=K] \,.
\label{eqn:leverage}
\end{equation}
We can solve for the leverage functions to write the SLV model in terms of the local volatilities, and we get
\begin{equation}
dF_t(T_i) = \eta_F(t,T_i,F_t(T_i)) \sqrt{ \dfrac{v_t}{\mathbb{E}[v_t|F_t(T_i)]} } \,dW^i_t \,.
\label{eqn:futures_master_equation}
\end{equation}

\begin{remark}
We notice that the conditions, imposed on the leverage functions so that the SLV dynamics has a solution, may be not respected by the particular form required by the Gy\"ongy Lemma. This is a known and open problem, see for instance \textcite{guyon2012being}, and its detailed analysis is beyond the scope of this paper.
\end{remark}

The problem of calibrating the leverage functions has been transformed into the simpler problem of calibrating the local volatilities $\eta_F$. This problem is considered and solved in a parsimonious way in \textcite{nastasi2020smile}. Here, we adopt such solution, that we briefly describe in the following paragraphs.

First, we introduce a common driving factor $s_t$ for all future prices. We can understand it as a normalised ``spot'' price. The dynamics of $s_t$ is given by
\begin{equation}
ds_t = a (1 - s_t) \,dt + s_t \eta(t,s_t) \,dW^s_t \,,
\end{equation}
where $a$ is the mean-reversion speed of the spot process, $\eta(t,K)$ is the local-volatility function of the spot price, and it is assumed to be Lipschitz, positive and bounded in price. Moreover, $W^s_t$ is a standard Brownian motion under the risk-neutral measure. Then, we assume that futures prices in the LV model can be calculated starting from the normalised spot price as
\begin{equation}
{\hat F}_t(T_i) := F_0(T_i) \,\mathbb{E}_t[s_{T_i}] = F_0(T_i) \left(1 - (1-s_t) \, e^{-a(T_i-t)}\right)
\label{eqn:original_master_relation}
\end{equation} 
for $i=1,\ldots,n$, where $F_0(T_i)$ is the term structure of futures prices observed in the market. Last equality in \eqref{eqn:original_master_relation} can be derived by straightforward algebra due to the particular form of the spot price dynamics. Furthermore, we can apply the It\^o Lemma and compare the result with equation \eqref{eqn:futures_master_equation}, so that we get
\begin{equation}
\eta_F(t,T_i,K) := \left( K - F_0(T_i) \left( 1 - e^{-a(T_i-t)} \right) \right) \eta(t,k_F(t,T_i,K)) \,,
\label{eq:etaF}
\end{equation}%
where the effective strike $k_F$ can be defined as
\begin{equation}
k_F(t,T_i,K) := 1 - e^{a(T_i-t)} \left( 1 - \frac{K}{F_0(T_i)} \right) \,.
\label{eq:effK}
\end{equation}%

Thus, the problem is now to calibrate the local volatility $\eta(t,K)$ of the spot price for a given mean-reversion speed $a$ to the prices of plain-vanilla options on futures. Notice that the problem is now much simpler since instead of calibrating a different function for each futures we need to calibrate only one function. In order to do so, we write the price $C(t,T_i,K)$ of plain-vanilla options with expiry $t$ on futures with maturity $T_i$ and strike $K$ in terms of the spot price (in order to simplify notation, we assume that expiry and payment of the options are on the same date):
\begin{equation}
C(t,T_i,K) := P_0(t) F_0(T_i) e^{-a(T_i-t)} c(t,k_F(t,T_i,K)) \,,
\label{eqn:call_price}
\end{equation}
where $P_0(t)$ is the zero-coupon bond with maturity $t$, and the normalized call prices $c(t,k)$ are defined as
\begin{equation}
c(t,k) := \mathbb{E}_t[(s_t-k)^+]
\end{equation}
and they satisfy the following extended version of the Dupire equation, see \textcite{nastasi2020smile}:
\begin{equation}
\partial _tc(t,k) = \left(-a-a(1-k)\partial_k+\dfrac{1}{2}k^2\eta^2 (t,k)\partial^2 _k\right)c(t,k) \,.
\label{eqn:plain_vanilla_PDE}
\end{equation}

Then, we can calibrate $\eta(t,k)$ by minimizing the distance between the market prices and the implied prices given by equation \eqref{eqn:call_price}. Notice that the procedure is very fast since at each step of the optimization we need to solve only once the PDE \eqref{eqn:plain_vanilla_PDE}. Once $\eta(t,k)$ is known we can calculate $\eta_F(t,T_i,K)$ by means of equation \eqref{eq:etaF}, and then we get the leverage functions $L(t,T_i,K)$ by solving equation \eqref{eqn:leverage}.

\subsection{The correlation structure}

Although the models introduced so far work for pricing plain-vanilla options on commodity indices in general, we are particularly interested in pricing S\&P GSCI ER indices. As discussed in Section \ref{sec:index}, in this specific case the rolling strategy depends only on futures contracts with maturities in two consecutive months. For this reason, the correlation structure can be only partially calibrated to market data. Indeed, the only parameters impacting the price of the index options are the entries immediately above and below the diagonal of the correlation matrix, namely the correlations between front and second futures contracts. This suggests that in the calibration procedure simulating all the different futures as different processes is not necessary, but only two process are required.

We re-write the correlation matrix to highlight the different roles of above and below diagonal terms w.r.t.\ the other entries:
\begin{equation}
\rho^i(t) := \sum_{j=1}^{n-i} \ind{t\in[T_j,T_{j+1})} \rho_{j,i+j} \,.
\label{eqn:rho}
\end{equation}
The function $\rho^i(t)$ for $i$ in $\{1,\ldots,n-1\}$ represents the correlation between two futures contracts whose maturities in the grid $\{T_1,\ldots,T_n\}$ are spaced by $i$ places. In particular $\rho(t) := \rho^1(t)$ is the correlation among consecutive futures contracts. Only this function can be calibrated with the given prices of index plain-vanilla options, while the other functions, with $1<i<n$, remain free parameters.

We can continue by extending the ideas of the previous section. Next, we introduce the two driving processes $s_t^c$ and $s_t^f$, that satisfy the SDEs:
\begin{equation}
ds^x_t = a (1-s^x_t) \,dt + s^x \eta(t,s^x_t) \,\sqrt{ \dfrac{v^x_t}{\mathbb{E}[v_t|s^x_t]} } \,dW^x_t
\;,\quad
x \in \{c,f\} \,,
\label{eqn:spot_fc}
\end{equation}
where $v^x_t$ has the same form as the variance process already defined. Moreover, we consider the same speed of mean-reversion $a$ and local-volatility function $\eta(t,k)$ entering the calibration of the normalized spot price process previously presented, while $W^c_t$ and $W^f_t$ are two standard Brownian motions under the risk-neutral measure with time-dependent correlation $\rho(t)$, i.e.,
\begin{equation}
d\langle W^c,W^f \rangle_t = \rho(t) \,dt \,.
\end{equation}

Then, we assume that the future prices ${\check F}_t(T_i)$ under this two-factor SLV model can be calculated as
\begin{equation}
{\check F}_t(T_i) := {\check F}^c_t(T_i) \ind{i {\rm~even}} + {\check F}^f_t(T_i) \ind{i {\rm~odd}},
\label{eqn:futures_check}
\end{equation}
where
\begin{equation}
{\check F}^x_t(T_i) := F_0(T_i) \,\mathbb{E}_t[s^x_{T_i}] = F_0(T_i) \left(1 - (1-s^x_t) \, e^{-a(T_i-t)}\right)
\;,\quad
x \in \{c,f\}.
\label{eqn:futures_sf}
\end{equation}

By direct calculation, we can prove that the above definition of future prices implies that for any $i$ in $\{1,\ldots,n-1\}$ the joint law of $({\check F}_t(T_i),{\check F}_t(T_{i+1}))$ is equivalent to the joint law of $(F_t(T_i),F_t(T_{i+1}))$, so that we can use the simpler two-factor SLV model to calibrate the correlation function $\rho(t)$ to index plain-vanilla options. First, we apply the It\^o's Lemma, and we get
\begin{equation}
d{\check F}^x_t(T_i) = \eta_F(t,T_i,{\check F}^x_t(T_i)) \,\sqrt{ \dfrac{v^x_t}{\mathbb{E}[v^x_t|{\check F}^x_t(T_i)]} } \,dW^x_t
\;,\quad
x \in \{c,f\},
\end{equation}
where the conditional expectation is now equivalently written w.r.t.\ the future price since ${\check F}^x_t(T_i)$ and $s^x_t$ are linked by a deterministic relationship. Then, we compare the above SDEs with the ones solved by futures price processes in the SLV model, see equation \eqref{eqn:futures_master_equation}. If we consider two consecutive futures contracts with maturities $T_i$ and $T_{i+1}$ (let us fix that $i$ is odd), the two SDEs satisfied by the price processes $F_t(T_i)$ and $F_t(T_{i+1})$ are the same SDEs satisfied by the price processes ${\check F}^c_t(T_i)$ and ${\check F}^f_t(T_i)$, when identifying the driving Brownian motions. Hence, the two pairs of processes have equivalent joint laws. Notice also the the marginal distributions of ${\check F}^x_t(T_i)$, ${\hat F}_t(T_i)$ and $F_t(T_i)$ are the same as expected.

At this point we can calibrate $\rho(t)$ by finding the two processes $s^c_t$ and $s^f_t$ which solve the two SDEs \eqref{eqn:spot_fc}, next calculating from their values the futures prices by means of equations \eqref{eqn:futures_check} and \eqref{eqn:futures_sf}, finally obtaining the index price by computing all the rolling periods as in equation \eqref{eq: evolution_roll}. Note that we need to use a Monte Carlo simulation to proceed. In the next section we describe the simulation scheme and the optimization strategy used to complete this goal.

\subsection{Simulation scheme}
\label{sec:simulation_scheme}


In order to calibrate plain-vanilla options on GSCI ER indices we need to resort to a Monte Carlo simulation since the rolling strategy \eqref{eq: evolution_roll} is too complex to be tackled with more efficient means. In the following sub-sections we describe the numerical scheme and the optimizer used to perform the calibration procedure.

Here, we describe the main steps of the Monte Carlo numerical discretization for the SLV model introduced in the previous sections. The dynamics we have selected for the variance process is of square-root type, see equation \eqref{eqn:variance}. We adopt this dynamics since it is the one used in the Heston model, and it is a common choice in SV models with good agreement with market behaviour.

The first difficulties in the simulation of equation \eqref{eqn:spot_fc} along with \eqref{eqn:variance} is the presence of the conditional expectation in the diffusive term. This kind of equations, named of Vlasov-McKean type, along with their role in the calibration of SLV models, are studied in the work of \textcite{guyon2012being}. 

The above scheme requires the evaluation of the conditional expectation at each time step, which in turn depends on the distribution of the processes $(s^x_t,v^x_t)$, that can be estimated by adopting a mean-field approximation. This means that we approximate such distribution with the empirical one by introducing $N$ interacting particles $\{(s^{x,i}_t,v^{x,i}_t)\}_{i=1,\ldots,N}$, such that
\begin{equation}
\frac{1}{N} \sum_{i=1}^N \delta(s-s^{x,i}_t)\delta(v-v^{x,i}_t) \xrightarrow[L^1, N\rightarrow \infty]{} p_{s^x_t,v^x_t}(s,v) \,.
\end{equation}
for any realization of the processes.

The above relationship, termed as propagation of chaos, allows us to use the Monte Carlo simulation of the interacting particles as a tool to approximate the following solution of the SLV SDEs:
\begin{equation}
ds^{x,i}_t = a (1-s^{x,i}_t) \,dt + s^{x,i}_t \eta(t,s^{x,i}_t) \,\sqrt{ \frac{ v^{x,i}_t \,\sum_{j=1}^N \delta(s^{x,i}_t-s^{x,j}_t) }{ \sum_{j=1}^N v^{x,j}_t \,\delta(s^{x,i}_t-s^{x,j}_t) } } \,dW^{x,i}_t
\end{equation}%
\begin{equation}
dv^{x,i}_t = \kappa (\theta - v^{x,i}_t) + \chi \sqrt{v^{x,i}_t} \,dW^{v,x,i}_t
\end{equation}%
for $i=0,\ldots,N$.

Now, before applying the standard Euler-Maruyama discretization scheme, we have to pay attention to the numerical issues coming from the possibility that the discretized variance process becomes negative. This situation may be exacerbated if the Feller condition is violated, a possibility occurring in practical calibration of square-root type models. In order to cure this problem we choose a full-truncated scheme as described in \textcite{lord2010comparison}, where the authors prove that the full-truncated scheme can achieve strong convergence to the continuous distribution for the Heston model. Furthermore, the scheme exhibits the smallest bias for preserving the positiveness of the variance process and it is comparable to an exact simulation scheme in terms of simulation error. Notice that in \textcite{tian2013hybrid} the full-truncated Euler-Maruyama approach is applied for an Heston SLV model, and it empirically seems to maintain the convergence properties already proved for the Heston model.

By following the full-truncated scheme we consider a time grid $\{t_k\}_{k=0,\ldots,M}$, and we denote by $s^{x,i}_{t_k}$ and $v^{x,i}_{t_k}$ the values of the discretized particle processes at time $t_i$ (with an abuse of notation we indicate the discretized processes as the continuous ones). Then, the scheme can be written as
\begin{equation}
s^{x,i}_{t_{k+1}} = s^{x,i}_{t_k} + a (1-s^{x,i}_{t_k}) \,\Delta t_k + s^{x,i}_{t_k} \eta(t,s^{x,i}_t) \,\sqrt{ \frac{ (v^{x,i}_{t_k})^+ \,\sum_{j=1}^N \delta^\epsilon(s^{x,i}_{t_k}-s^{x,j}_{t_k}) }{ \sum_{j=1}^N (v^{x,j}_{t_k})^+ \,\delta^\epsilon(s^{x,i}_{t_k}-s^{x,j}_{t_k}) } } \,\Delta W^{x,i}_{t_k},
\label{eqn:snum}
\end{equation}%
\begin{equation}
v^{x,i}_{t_{k+1}} = v^{x,i}_{t_k} + \kappa ( \theta - (v^{x,i}_{t_k})^+ ) \,\Delta t_k + \chi \sqrt{ ( v^{x,i}_{t_k} )^+ } \,\Delta W^{v,x,i}_{t_k},
\label{eqn:vnum}
\end{equation}%
for $i=0,\ldots,N$ and $k=0,\ldots,M-1$, where $s^{x,i}_0 = 1$ and $v^{x,i}_0 = v_0$. The function $\delta^\epsilon(x)$ is any mollifier of the Dirac delta, for instance it can be defined as a Gaussian density centered in zero with standard deviation $\epsilon$. The local volatility $\eta(t,k)$ is obtained by making a piecewise-constant interpolation in time and a piecewise linear interpolation in price. 
 
\subsection{Hybrid global-local model calibration}\label{sec:global_local_calibration}

 So far we have calibrated the leverage function to vanilla options on futures using Gy\"ongy's lemma. Moreover, we argued that only the correlation between consecutive futures can have an impact on the prices of plain-vanilla options on GSCI ER indices. Therefore, we can define the vector $\mathbf{p}$ containing the remaining six degrees of freedom in the model:
\begin{equation}
\mathbf{p} := \{a,\kappa,\theta,\chi,\rho_v,v_0,\rho\} \,,
\end{equation}
where the last component is the correlation function defined in \eqref{eqn:rho} in terms of the front-second futures correlations.\\
Next, our goal is to fit these remaining degrees of freedom so that our model can recover the market prices of plain-vanilla options on GSCI ER indices. For this purpose, we start by defining a metric $f_p$ to measure how far the model prices are from the market prices of plain-vanillas on the index. More precisely, we choose the function $f_p$ as the $p$-norm of the vector containing the difference between the market prices $I_{\rm market}$ and the model prices $I_{\rm model}(\mathbf{p})$ of plain vanillas on the index, the latter depending on the model parameters in vector $\mathbf{p}$, i.e.:
\begin{equation}\label{eqn:loss_function}
    f_p(\mathbf{p}) = \left(\sum _{j = 1}^J\left|I^j_{\rm market}-I^j_{\rm model}(\mathbf{p})\right|^p \right)^{\dfrac{1}{p}}\,,
\end{equation}
where $J$ is the total number of vanillas on the index that have been considered.\\

In this setup, the calibration of the free parameters to fit the prices of plain-vanillas on the index can be formulated as the following unconstrained global optimization problem in a bounded domain:
\begin{equation}\label{eqn:minimization_problem}
    \min_{\mathbf{p}\in D\subseteq \mathbb{R}^n}f_p(\mathbf{p}) \,,
\end{equation}
where $f_p$ is the cost function defined on $D = \Pi_{i = 1}^n[l_i,u_i]$, with $l_i$ and $u_i$ being the lower and upper bounds in direction $i$, respectively. The solution vector $\mathbf{p^*}$ contains the calibrated parameters and is defined as:
\begin{equation}\label{eqn:parameters_solution}
        \mathbf{p^*} = \argmin{{\mathbf{p}\in D\subseteq \mathbb{R}^n}}f_p(\mathbf{p}),
\end{equation}
Note that the cost function depends on $I_{\rm model}(\mathbf{p})$, so that each single evaluation of the cost function requires the numerical solution of the model for a given set of parameters, that is, performing a Monte Carlo simulation with these fixed parameters. \\

In order to solve the unconstrained global optimization problem (\ref{eqn:minimization_problem}), we propose a simplified version of a two-phase calibration strategy (see \textcite{FERREIROFERREIRO2020467, two_phase} for a complete description). In particular, we start by defining an initial guess $\mathbf{p}_0$ of the optimal solution $\mathbf{p}^*$. We typically choose $\mathbf{p}_0$ randomly. Next, we run a global optimization algorithm starting with $\mathbf{p}_0$. Once the global algorithm has finished, it provides a new vector $\mathbf{p}_1$ which gives an approximation of $\mathbf{p}^*$. Then, we use $\mathbf{p}_1$ as the initial point for a local optimization algorithm. Finally, the local optimization algorithm gives us $\mathbf{p}_2$. These steps are summarized in Algorithm \ref{alg:overall}. 

\begin{algorithm}[H]
\caption{Overall optimization algorithm}\label{alg:overall}
\begin{algorithmic}
\State \textbf{Input:}
\Indent
\State $\mathbf{p}_0$ \Comment{random seed for $\mathbf{p}$}
\State $f$ \Comment{function to be minimized}
\EndIndent
\State \textbf{Output:}
\Indent
\State $\mathbf{p}$ \Comment{approximated value of $\mathbf{p^*}$}
\EndIndent
\State \textbf{Algorithm:}
\Indent
\State $\mathbf{p}_1 = \text{ESCH}(\mathbf{p}_1)$ \Comment{global minimization.}
\State $\mathbf{p}_2 = \text{Subplex}(\mathbf{p}_1)$ \Comment{local minimization.}
\EndIndent
\end{algorithmic}
\end{algorithm}
As global optimization algorithm, in this work we use the so called ESCH (see \textcite{DE_variation}). ESCH algorithm belongs to the broad category of Evolutionary Algorithms (EAs) (see \textcite{EA}) which is a class of heuristics inspired by natural selection in biological populations. We specifically use ESCH algorithm because it is open access available in the \textit{NLopt nonlinear-optimization package}\footnote{The NLopt nonlinear-optimization package is one of the de facto standards for non-linear optimization in various programming languages such as \textit{C++}, \textit{Julia}, \textit{Rust}... The library can be found in \href{shttp://github.com/stevengj/nlopt}{http://github.com/stevengj/nlopt} and is developed and mantained by Steven G. Johnson}. In Algorithm \ref{alg:differential_evolution} we show the pseudocode for the ESCH method.
This evolutionary algorithm starts by dividing the population into two groups: parents and offspring. First, the parameters of the parents are randomly initialized from a uniform distribution. Then, some of the parents are selected and their information recombined to generate offspring. Some of the offspring would mutate one of their parameters. Next, we assign a score to the offspring based on their value in the function that we want to minimize. Finally, all individuals are ranked according to their fit score, the less fit individuals are
removed from the population and only the best individuals are
stored along the generations. The adopted recombination and
the mutation operators are the so-called single point and the
Cauchy distribution, respectively. The scheme for the evolution can be found in Algorithm \ref{alg:differential_evolution}. For a more detailed description we refer to the original thesis \textcite{subplex}.

The advantage of starting by a global optimization algorithm is that it is able to escape from local minima. However, in general these algorithms in are computationally very expensive as they have a slow rate of convergence. Our purpose starting with a global optimization algorithm is to explore the space and end up with a good initial point for the local optimization algorithm. In the best case scenario the output of the algorithm should lay in the convex region defined by the global minima. Other possible choices for the global optimization routine are: Simulated Annealing (SA, see \textcite{EA}, \textcite{aarts1985statistical}), Differential Evolution (DE, see \textcite{storn1997differential}) or Particle Swarm (PS, see \textcite{kennedy1995particle}).
 
\begin{algorithm}[H]
\caption{ESCH pseudocode}\label{alg:differential_evolution}
\begin{algorithmic}
\State \textbf{Input:}
\Indent
\State $n$ \Comment{problem dimension}
\State $f$ \Comment{function to be minimized}
\State $x$ \Comment{initial approximation to minimum}
\State $np$ \Comment{number of parents}
\State $no$ \Comment{number of offspring}
\EndIndent
\State \textbf{Output:}
\Indent
\State $x$ \Comment{computed minimum}
\EndIndent
\State \textbf{Algorithm:}
\Indent
\State Initialize parents and offspring population
\State Parents fitness evaluation
\While{termination test not satisfied}
\State Crossover
\State Gaussian Mutation
\State Offspring fitness evaluation
\State Selection of the fittest as parents
\EndWhile
\EndIndent
\end{algorithmic}
\end{algorithm}
 As local optimization algorithm, in this article we use a variation of the Nelder-Mead algorithm (see \textcite{Nelder1965ASM}), called ``Subplex'' and developed in \textcite{subplex}. The Subplex algorithm starts by dividing the search space into subspaces. Then, the Nelder-Mead algorithm is used to minimize in each subspace. The subspaces in which the minimization has been larger are joined to form a new subspace. The process continues iteratively until the stopping criteria is met. An outline of the Subplex basic steps can be found in Algorithm \ref{alg:subplex}. For a more detailed description we refer to the original thesis \textcite{subplex}.
Note that local search algorithms are not able to escape from local minima, although they exhibit faster convergence rates than the global optimization ones. Therefore, the purpose of using the local optimization algorithm is to perform a fine grain optimization much faster than it would be possible with a global optimization method.

\begin{algorithm}[H]
\caption{Subplex pseudocode}\label{alg:subplex}
\begin{algorithmic}
\State \textbf{Input:}
\Indent
\State $n$ \Comment{problem dimension}
\State $f$ \Comment{function to be minimized}
\State $x$ \Comment{initial approximation to minimum}
\State $Scale$ \Comment{initial step size for the $n$ coordinate directions}
\State $\alpha$ \Comment{reflection coefficient}
\State $\beta$ \Comment{contraction coefficient}
\State $\gamma$ \Comment{expansion coefficient}
\State $\delta$ \Comment{shrinkage coefficient}
\State $\psi$ \Comment{simplex reduction coefficient}
\State $\Omega$ \Comment{step reduction coefficient}
\State $nsmin$ \Comment{minimum subspace dimension}
\State $nsmax$ \Comment{maximum subspace dimension}
\EndIndent
\State \textbf{Output:}
\Indent
\State $x$ \Comment{computed minimum}
\EndIndent
\State \textbf{Algorithm:}
\Indent
\While{termination test not satisfied}
\State Set stepsizes
\State Set subspaces
\For{each subspace}
\State Use Nelder Mead Simplex to search subspace
\State Check Termination
\EndFor
\EndWhile
\EndIndent
\end{algorithmic}
\end{algorithm}

\section{Numerical investigations}
\label{sec:numerics}


In this section we present the numerical experiments to validate the proposed model against real market data. In particular we focus on the S\&P GSCI Crude Oil ER index.

\subsection{Sensitivity of index options prices with respect to parameters}
As we do not expect that all the parameters have an impact on the prices of the plain vanillas on the S\&P GSCI Crude Oil ER index, in this section we explore which of them play a sensitive role in their pricing. For this purpose, we will fix a set of reference parameters (see Table \ref{tab:base_parameters}) and vary one parameter at a time keeping the rest constant to evaluate its impact on the prices of plain vanillas on the index. The baseline parameters are obtained from the calibration of the model to real market data performed in Section \ref{sec:model_calibration}. The variations over the reference parameters are extreme values to clearly see the potential impact that each of the parameters has.
\begin{table}[H]
    \centering
    \begin{tabular}{|c|c|c|}
        \hline
         \textbf{Name} & \textbf{Symbol} & \textbf{Value} \\
         \hline
         Mean reversion speed & $a$ & $0.3$ \\
         \hline
         Asset Correlation & $\rho$ & $0.9$\\
         \hline
         Stochastic volatility mean reversion speed & $\kappa$ &  $1.0$\\
         \hline
         Long term level & $\theta$ & $1.0$\\
         \hline
         Vol-of-vol & $\chi$ & $0.1$\\
         \hline
         Stochastic Volatility correlation & $\rho_v$ & $0.0$\\
         \hline
         Initial stochastic volatility & $v_0$  & $1.0$\\
         \hline
    \end{tabular}
    \caption{Set of reference parameters.}\label{tab:base_parameters}
    \label{tab:reference_values}
\end{table}
This section is divided in two. In Section \ref{sec:time_structure} we show the impact of the parameters for different maturities. In Section \ref{sec:smiles} we show the impact of the parameters for different strikes.
\subsubsection{Time structure}\label{sec:time_structure}
For short maturities, the prices of plain vanillas on the index should not depend heavily on any parameters apart from the local volatility. This is roughly motivated by the fact that investing on the index reassembles an investment on the front month futures contract. Hence, in the limit case where the maturity is extremely short and there has not been any rolling period, we shouldn't be able to distinguish between investing in an option on the index or investing on the futures contract itself. From this reasoning it follows that, as the prices of a plain vanillas do not depend on the path but on the terminal distribution, the only relevant parameter for pricing them is the local volatility.

In order to avoid possible misleadings, note that the path dependence of the plain vanilla on an index comes from the definition of the index with respect to the futures and not from the payoff of the contract. From the point of view of the index process, the prices of the plain vanillas just depend on the distribution at maturity and not on the path of the index process. From the point of view of the futures processes, the prices of the plain vanillas depend on the their path. As a result, it is only for longer maturities where we will be able to see a clear impact of all the parameters different from the local volatility. 

\begin{figure}[H]
    \centering
    \begin{tikzpicture}
    \begin{axis}[
    width = 0.5\linewidth,
    xmajorgrids=true,
    ymajorgrids=true,
    legend pos=north west ,
    ymin=0.24,
    ymax=0.35,
    ]
\addplot[
    color=black,
    mark=square
    ]%
table[
    x=t,
    y=y
    ]{data/atm_a=0.dat};
\addlegendentry[]{$a=0$}
\addplot[
    color=black,
    mark=*
    ]%
table[
    x=t,
    y=y
    ]{data/atm_a=1.dat};
\addlegendentry{$a=1$}
\end{axis}
\end{tikzpicture}
    \begin{tikzpicture}
    \begin{axis}[
 width = 0.5\linewidth,
    xmajorgrids=true,
    ymajorgrids=true,
    legend pos=south east,
    ymin=0.24,
    ymax=0.35,
    ]
\addplot[
    color=black,
    mark=square
    ]%
table[
    x=t,
    y=y
    ]{data/atm_rho=-1.dat};
\addlegendentry{$\rho=-1$}
\addplot[
    color=black,
    mark=*
    ]%
table[
    x=t,
    y=y
    ]{data/atm_rho=1.dat};
\addlegendentry{$\rho=1$}
\end{axis}
\end{tikzpicture}\\
    \begin{tikzpicture}
    \begin{axis}[
    width = 0.5\linewidth,
    xmajorgrids=true,
    ymajorgrids=true,
    legend pos=south east ,
    ymin=0.24,
    ymax=0.35,
    ]
\addplot[
    color=black,
    mark=square
    ]%
table[
    x=t,
    y=y
    ]{data/atm_rhov=-1.dat};
\addlegendentry{$\rho_v=-1$}
\addplot[
    color=black,
    mark=*
    ]%
table[
    x=t,
    y=y
    ]{data/atm_rhov=1.dat};
\addlegendentry{$\rho_v=1$}
\end{axis}
\end{tikzpicture}
    \begin{tikzpicture}
    \begin{axis}[
    width = 0.5\linewidth,
    xmajorgrids=true,
    ymajorgrids=true,
    legend pos=north east ,
    ymin=0.24,
    ymax=0.35,
    ]
\addplot[
    color=black,
    mark=square
    ]%
table[
    x=t,
    y=y
    ]{data/atm_chi=0.dat};
\addlegendentry{$\chi=0$}
\addplot[
    color=black,
    mark=*
    ]%
table[
    x=t,
    y=y
    ]{data/atm_chi=1.dat};
\addlegendentry{$\chi=1$}
\end{axis}
\end{tikzpicture}
    \caption{Prices of the plain vanillas on the S\&P GSCI Crude Oil ER index for different variations of the reference parameters from Table \ref{tab:base_parameters}. The $x$ axis represents the maturity in months and the $y$ axis implied volatility. The data for the simulation is taken from the future prices of WTI Crude oil on the 16 December 2019.}
    \label{fig:time_structure_relevant}
\end{figure}
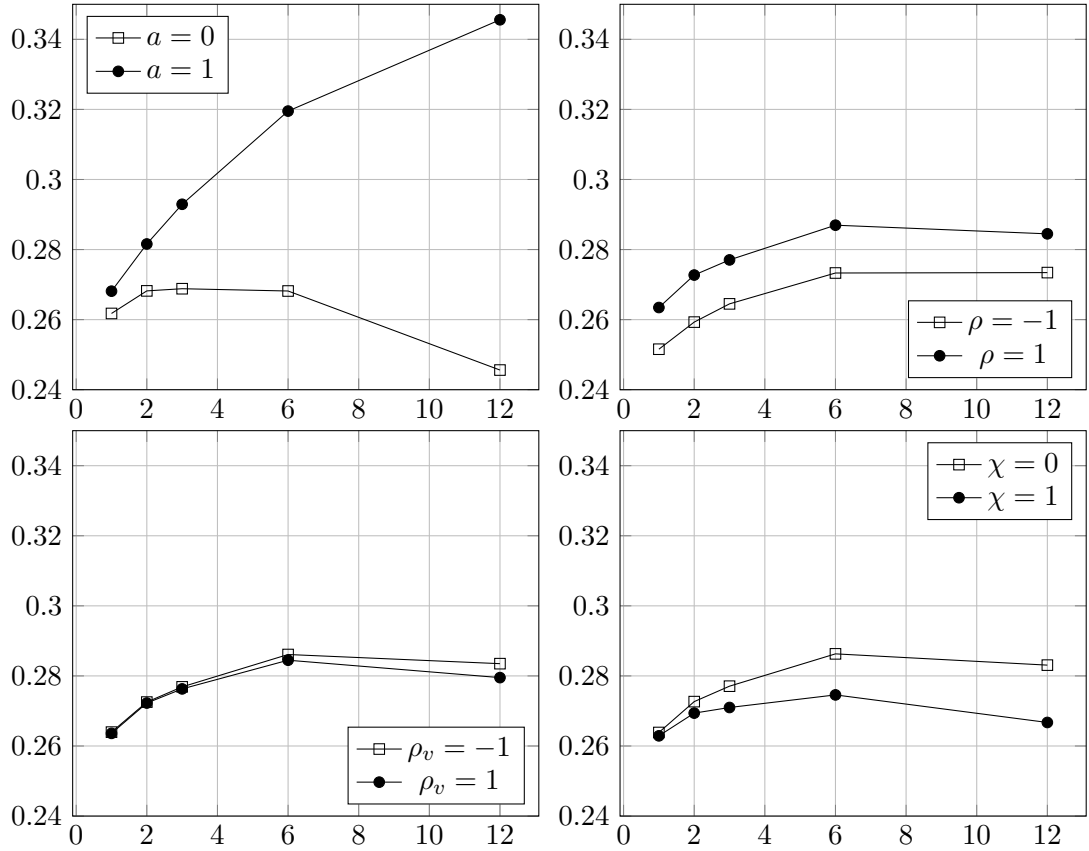

In Figure \ref{fig:time_structure_relevant} we depict the implied volatilities of the ATM calls on the index for different maturities and variations of $a$, $\rho$, $\rho_v$ and $\chi$ from the reference parameters. Approximately we can qualitatively distinguish two effects. On the one hand, $a$ and $\chi$ produce an increasing impact on long maturities. On the other hand, $\rho$ changes the level from the first maturity, and its effect does not seem to increase over time.

The impact of $\rho_v$, $\kappa$, $\theta$ and $v_0$  in the implied volatilities of the ATM calls on the index for different maturities is marginal compared with the effect of the other parameters.

\subsubsection{Smiles}\label{sec:smiles}
As argued in Section \ref{sec:time_structure}, the effect of the set of parameters $\mathbf{p}$ is more relevant for longer maturities. For that reason, in this section we will analyze the effect of varying the parameters $\mathbf{p}$ in the smile for the one year maturity call.

\begin{figure}[H]
\centering
\begin{tikzpicture}
\begin{axis}[
    width = 0.5\linewidth,
    xmajorgrids=true,
    ymajorgrids=true,
    legend pos=north east ,
    ymin=0.15,
    ymax=0.45
    ]
\addplot[
    color=black,
    mark=square
    ]%
table[
    x=moneyness,
    y=y
    ]{data/time12_a=0.dat};
\addlegendentry{$a=0$}
\addplot[
    color=black,
    mark=*
    ]%
table[
    x=moneyness,
    y=y
    ]{data/time12_a=1.dat};
\addlegendentry{$a=1$}
\end{axis}
\end{tikzpicture}
\begin{tikzpicture}
\begin{axis}[
    width = 0.5\linewidth,
    xmajorgrids=true,
    ymajorgrids=true,
    legend pos=north east ,
    ymin=0.15,
    ymax=0.45
    ]
\addplot[
    color=black,
    mark=square
    ]%
table[
    x=moneyness,
    y=y
    ]{data/time12_rho=-1.dat};
\addlegendentry{$\rho=-1$}
\addplot[
    color=black,
    mark=*
    ]%
table[
    x=moneyness,
    y=y
    ]{data/time12_rho=1.dat};
\addlegendentry{$\rho=1$}
\end{axis}
\end{tikzpicture}\\
\begin{tikzpicture}
\begin{axis}[
    width = 0.5\linewidth,
    xmajorgrids=true,
    ymajorgrids=true,
    legend pos=north east ,
    ymin=0.15,
    ymax=0.45
    ]
\addplot[
    color=black,
    mark=square
    ]%
table[
    x=moneyness,
    y=y
    ]{data/time12_rhov=-1.dat};
\addlegendentry{$\rho_v=-1$}
\addplot[
    color=black,
    mark=*
    ]%
table[
    x=moneyness,
    y=y
    ]{data/time12_rhov=1.dat};
\addlegendentry{$\rho_v=1$}
\end{axis}
\end{tikzpicture}
\begin{tikzpicture}
\begin{axis}[
    width = 0.5\linewidth,
    xmajorgrids=true,
    ymajorgrids=true,
    legend pos=north east ,
    ymin=0.15,
    ymax=0.45
    ]
\addplot[
    color=black,
    mark=square
    ]%
table[
    x=moneyness,
    y=y
    ]{data/time12_chi=0.dat};
\addlegendentry{$\chi=0$}
\addplot[
    color=black,
    mark=*
    ]%
table[
    x=moneyness,
    y=y
    ]{data/time12_chi=1.dat};
\addlegendentry{$\chi=1$}
\end{axis}
\end{tikzpicture}
    \caption{Prices of the one year plain vanillas on the S\&P GSCI Crude Oil ER index for different variations of the reference parameters from Table \ref{tab:base_parameters}. The $x$ axis represents the moneyness and the $y$ axis implied volatility. The data for the simulation is taken from the future prices of WTI Crude oil on the 16 December 2019.}
    \label{fig:smile_structure_relevant}
\end{figure}
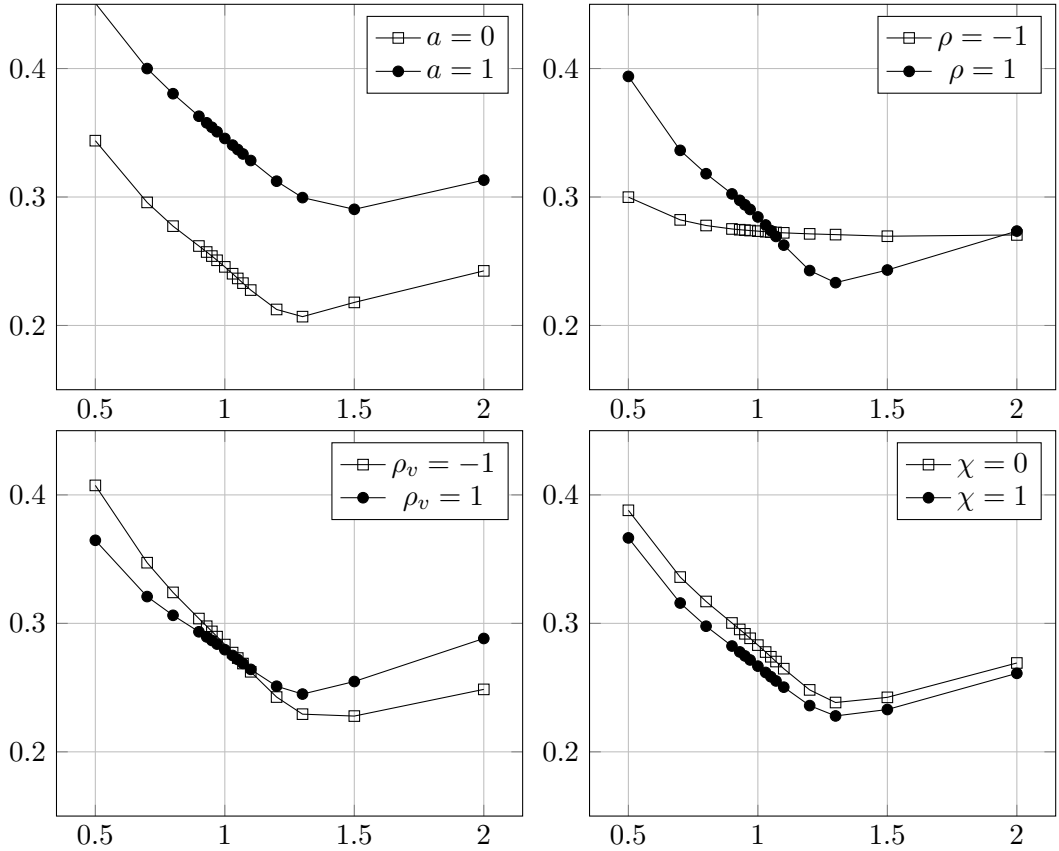

In Figure \ref{fig:smile_structure_relevant} we depict the implied volatilities of one year maturity calls on the index for different moneyness and variations of $a$, $\rho$, $\rho_v$ and $\chi$ from the reference parameters. Once again we can distinguish two different behaviours. First, $a$ seems to impact significantly the level of the smile, while the impact of $\chi$ is much smaller. Second, $\rho$ and $\rho_v$ impact the shape of the smile by flattening or rotating it.

The impact of $\kappa$, $\theta$ and $v_0$  in the implied volatilities of the one year maturity calls on the index for different moneyness is again negligible compared with the effect of the other parameters.

\subsection{Model calibration}\label{sec:model_calibration}
In the previous section we illustrated that not all the parameters have a relevant impact on the prices of plain vanillas on the index. The reason for this is that, the impact of these parameters is counterbalanced by the calibration of the local volatility term, so that there is not effect on index option prices. A similar situation arises in other SLV models commonly used in the practice, see for instance \textcite{impact_parameters}.
Hence, we can set some reasonable values a-priori for those parameters with less effect on the price. In particular, we constrain the original parameter vector $\mathbf{p}$ to
\begin{equation*}
    \mathbf{\hat{p}} := \{a,\chi,\rho_v,\rho\}.
\end{equation*}
For the other parameters we will consider from now on the fixed values appearing in Table 2.
\begin{table}[H]
    \centering
    \begin{tabular}{|c|c|c|}
        \hline
         \textbf{Name} & \textbf{Symbol} & \textbf{Calibrated Value}  \\
         \hline
         Stochastic volatility mean reversion speed & $\kappa$ & $1.0$\\
         \hline
         Long term level & $\theta$ & $1.0$\\
         \hline
         Initial volatility & $v_0$ & $1.0$\\
         \hline
    \end{tabular}
    \caption{Fixed values of the parameters not affecting the price.}
    \label{tab:fixed_values}
\end{table}
Analogously, the previous unconstrained global optimization problem in a bounded domain defined in Equation \eqref{eqn:minimization_problem} is reduced to to:
\begin{equation}\label{eqn:reduced_minimization_problem}
    \min_{\mathbf{\hat{p}}\in \hat{D}\subseteq \mathbb{R}^{\hat{n}}}f_p(\mathbf{\hat{p}}) \,,
\end{equation}
where $f_p$ is defined on $\hat{D} = \Pi_{i = 1}^{\hat{n}}[l_i,u_i]$, with $l_i$ and $u_i$ being the lower and upper bounds in direction $i$, respectively.

Market quotes for plain vanilla options on GSCI indices are not available in the market. However, consensus quotes can be obtained from IHS Markit Totem service on the last day of each month. Specifically we have consensus data for $30$ November $2019$ and $31$ December $2019$. On the contrary, the data we had available for the future prices and the prices of the plain vanillas on futures are from the $16$ December $2019$. As we cannot directly calibrate against market quotes of the same day for the futures, plain vanillas on futures and plain vanillas on the index, we have to slightly modify the cost function to account for this matter. The proposed modification consists in computing the $p$-norm between the model price and the mean market price of November and December and adding a normalization factor to penalise more the points laying outside the market range and less the ones laying inside the market range:
\begin{equation}\label{eqn:normalised_loss_function}
    \hat{f}_p(\mathbf{p}) = \left(\sum _{j = 1}^J\dfrac{\left|I^j_{\rm mean}-I^j_{\rm model}(\mathbf{p})\right|^p}{\left|I^j_{\rm December}-I^j_{\rm November}(\mathbf{p})\right|^p} \right)^{\dfrac{1}{p}}\,,
\end{equation}
Under all these considerations, we obtain the values in Table 3 for the parameters in vector $\hat{p}$ after calibrating the model to the vanillas on the index.
\begin{table}[H]
    \centering
    \begin{tabular}{|c|c|c|c|}
        \hline
         \textbf{Name} & \textbf{Symbol} & \textbf{Seed} & \textbf{Calibrated Value}  \\
         \hline
         Mean reversion speed & $a$ & $0.1$ & $0.267419$\\
         \hline
         Asset Correlation & $\rho$ & $0.0$ & $0.86381$\\
         \hline
         Vol-of-vol & $\chi$ & $1.0$ & $0.0287296$\\
         \hline
         Stochastic Volatility correlation & $\rho_v$ & $1.0$ & $-0.18058$\\
         \hline
    \end{tabular}
    \caption{Calibrated values of $\hat{\mathbf{p}}$ against the S\&P GSCI Crude Oil ER index 1 Year Call Options quoted on 30 November 2019 and the S\&P GSCI Crude Oil ER index 1 Year Call Options quoted on 31 December 2019, both on CME market. For this calibration the hybrid global-local procedure described in Section \ref{sec:global_local_calibration} has been used.}\label{tab:calibrated_values}
\end{table}
If we represent the results obtained by the model with the calibrated parameters of Table \ref{tab:calibrated_values} we obtain the following results.
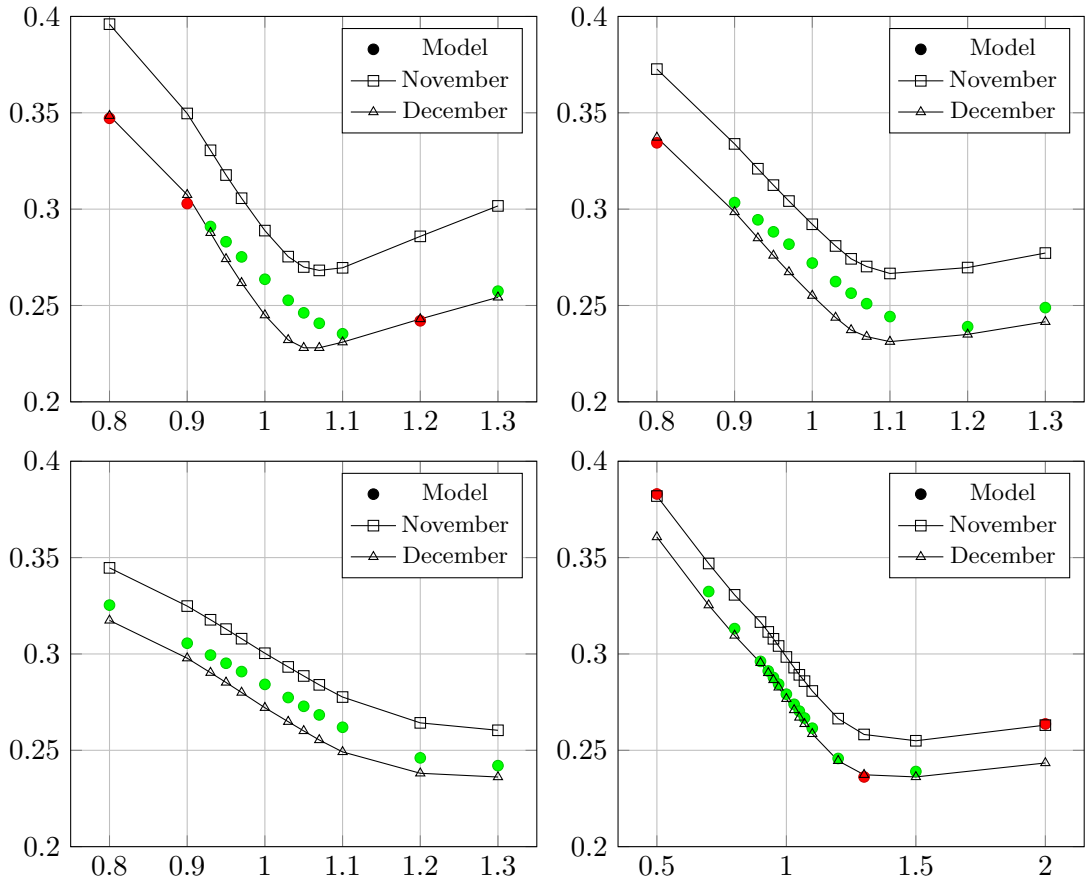
\begin{figure}[H]
\centering
\begin{tikzpicture}
\pgfplotsset{
    colormap={binary}{
        color=(red) color=(green)
    }
}
\begin{axis}[
    width = 0.5\linewidth,
    xmajorgrids=true,
    ymajorgrids=true,
    legend pos=north east ,
    legend style={font=\footnotesize},
    ymin=0.20,
    ymax=0.4
]
\addplot[
    scatter,
    only marks,
    scatter src = explicit,
    ]%
table[
    x=moneyness,
    y=y,
    meta=color
    ]{data/calibrated_time1.dat};
\addlegendentry{Model}
\addplot[
    color=black,
    mark=square
    ]%
table[
    x=moneyness,
    y=y
    ]{data/time1_November.dat};
\addlegendentry{November}
\addplot[
    color=black,
    mark=triangle
    ]%
table[
    x=moneyness,
    y=y
    ]{data/time1_December.dat};
\addlegendentry{December}
\end{axis}
\end{tikzpicture}
\begin{tikzpicture}
\pgfplotsset{
    colormap={binary}{
        color=(red) color=(green)
    }
}
\begin{axis}[
    width = 0.5\linewidth,
    xmajorgrids=true,
    ymajorgrids=true,
    legend pos=north east ,
    legend style={font=\footnotesize},
    ymin=0.20,
    ymax=0.4
]
\addplot[
    scatter,
    only marks,
    scatter src = explicit,
    ]%
table[
    x=moneyness,
    y=y,
    meta=color
    ]{data/calibrated_time2.dat};
\addlegendentry{Model}
\addplot[
    color=black,
    mark=square
    ]%
table[
    x=moneyness,
    y=y
    ]{data/time2_November.dat};
\addlegendentry{November}
\addplot[
    color=black,
    mark=triangle
    ]%
table[
    x=moneyness,
    y=y
    ]{data/time2_December.dat};
\addlegendentry{December}
\end{axis}
\end{tikzpicture}
\begin{tikzpicture}
\pgfplotsset{
    colormap={binary}{
        color=(green) color=(red)
    }
}
\begin{axis}[
    width = 0.5\linewidth,
    xmajorgrids=true,
    ymajorgrids=true,
    legend pos=north east ,
    legend style={font=\footnotesize},
    ymin=0.20,
    ymax=0.4
]
\addplot[
    scatter,
    only marks,
    scatter src = explicit,
    ]%
table[
    x=moneyness,
    y=y,
    meta=color
    ]{data/calibrated_time6.dat};
\addlegendentry{Model}
\addplot[
    color=black,
    mark=square
    ]%
table[
    x=moneyness,
    y=y
    ]{data/time6_November.dat};
\addlegendentry{November}
\addplot[
    color=black,
    mark=triangle
    ]%
table[
    x=moneyness,
    y=y
    ]{data/time6_December.dat};
\addlegendentry{December}
\end{axis}
\end{tikzpicture}
\begin{tikzpicture}
\pgfplotsset{
    colormap={binary}{
        color=(red) color=(green) 
    }
}
\begin{axis}[
    width = 0.5\linewidth,
    xmajorgrids=true,
    ymajorgrids=true,
    legend pos=north east ,
    legend style={font=\footnotesize},
    ymin=0.20,
    ymax=0.4
]
\addplot[
    scatter,
    only marks,
    scatter src = explicit,
    ]%
table[
    x=moneyness,
    y=y,
    meta=color
    ]{data/calibrated_time12.dat};
\addlegendentry{Model}
\addplot[
    color=black,
    mark=square
    ]%
table[
    x=moneyness,
    y=y
    ]{data/time12_November.dat};
\addlegendentry{November}
\addplot[
    color=black,
    mark=triangle
    ]%
table[
    x=moneyness,
    y=y
    ]{data/time12_December.dat};
\addlegendentry{December}
\end{axis}
\end{tikzpicture}

    \caption{The line with squares shows the S\&P GSCI Crude Oil ER index 1 Year Call Options for different moneyness quoted on 30 November 2019 on CME
market. The line with triangles shows the S\&P GSCI Crude Oil ER index 1 Year Call Options for different moneyness quoted on 31 December 2019 on CME
market. The dots represent the model prediction for the S\&P GSCI Crude Oil ER index 1 Year Call Options for different moneyness with the computation started on the 16 December 2019. Green color indicates that the model price lays between the market prices, otherwise red color is used.}
    \label{fig:calibrated}
\end{figure}
\subsection{Computer implementation details}
Concerning the hardware configuration, all tests have been performed in a Ubuntu server running over a virtualization layer (VMware) with 8 GB of RAM, 16 CPU cores (Intel(R) Xeon(R) CPU E5-2650 v4 at 2.2GHz for a total of 16 logical threads).

For the software, we have done the implementation in the \textit{C}{\tiny++} programming language using the OpenMP library to take advantage of the multithreading capabilities of the processor. The GNU $C${\tiny++} compiler has been used \footnote{Also the Clang and the Intel C++ compilers have been tested obtaining the same or very similar results.}.

With respect to execution time, a single evaluation of the cost function (computing the implied volatilities of the plain vanillas on the index) takes around $7$ seconds. The whole calibration algorithm requires roughly in two hours. Note that this whole calibration algorithm is normally executed only the last day of each month when the consensus quotes on vanilla options on the index are obtained. Assuming small variations of the future prices on any other date different from the last day of month, we can avoid running the whole calibration algorithm. Instead we can just run the local calibration algorithm taking the parameters of the previous day as initial seed. In this scenario the computational time of the calibration could be reduced significantly to about half an hour. Therefore, these computational times are reasonable for the real practice of the markets. Indeed, if we consider that the calibration can be run as an overnight process.
\section{Conclusions and further research}
In the present article we have presented a new stochastic local volatility model for the pricing of derivative contracts on commodity futures. Moreover, a calibration methodology is proposed, which requires the use of hybrid global-local optimization algorithms.

As illustrated in Section \ref{sec:numerics}, the model captures the essential behaviour of plain vanillas on the index. This opens different possibilities. The most immediate one is to use the model to build the next consensus. The consensus is built from the data provided by different financial institutions. Those financial institutions providing data that differs significantly from the consensus are expelled from the consensus data source.

Another application of the proposed model would be to price exotics on the index. For this task, usually a model on the index is directly considered and its calibration has to be performed directly on the exotics. The problem with this approach is that data of exotics on the index is scarce. For this reason the calibration of models built directly upon the index is difficult. With the proposed model we have already obtained a calibration of the different parameters and we can directly use it to price exotics. 

Among the possible future research lines, we specifically mention two interesting ones. On the one hand, the methodology here described could be extended to price other indexes, presumably more complex ones. On the other hand, we could explore alternative models for the stochastic volatility. A possibility comes from extending the Heston model to admit powers of the stochastic volatility in the diffusion term of the stochastic volatility dynamics. This extension would be motivated by the fact that none of the parameters $(\kappa,\theta,v_0)$ of the stochastic volatility affected the prices of the plain vanillas on the index. This suggests that a further modifications of the drift term will not have any impact on the prices of plain vanillas. However, the diffusion term had an impact. Therefore, by considering powers of the volatility in the diffusion term of the stochastic volatility dynamics we could enhance the calibration of the plain vanilla options on the index.

\section*{Acknowledgements} A.P. Manzano-Herrero and C. V\'azquez acknowledge the support received from the Centro de Investigaci\'on en Tecnolog\'{\i}as de la Informaci\'on y las Comunicaciones de Galicia, CITIC, funded by Xunta de Galicia and the European Union (European Regional Development Fund, Galicia 2014-2020 Program) by grant ED431G 2019/01. Also both authors acknowledge the funding from Xunta de Galicia through the grant ED431C2018/033, as well as the funding from Spanish Ministry of Science and Innovation with the grant PID2019-10858RB-I00. 

\printbibliography

\end{document}